\documentclass[prl,amsmath,amssymb,twocolumn,superscriptaddress,floatfix]{revtex4}
\usepackage{bbm}
\usepackage{graphicx}
\usepackage{dcolumn}
\usepackage{bm}
\usepackage{subfigure}
\usepackage{amsmath}
\usepackage{feynmf}
\usepackage{hyperref}

\usepackage{attachfile}
\newcommand{\ua}{\uparrow}
\newcommand{\da}{\downarrow}
\newcommand{\ra}{\rightarrow}

\newcommand{\SRO}{Sr$_2$RuO$_4$}
\newcommand{\bk}{\mathbf k}

\usepackage{times}

\begin{document}
\title{Possible `symmetry-imposed' near-nodal two-dimensional p-wave pairing in \SRO}
\author{Yu Li}
\affiliation{Shenzhen Institute for Quantum Science and Engineering and Department of Physics, Southern University of Science and Technology, Shenzhen 518055, China}
\affiliation{Center for Quantum Computing, Peng Cheng Laboratory, Shenzhen 518005, China}
\author{Wen Huang}
\email{huangw3@sustech.edu.cn}
\affiliation{Shenzhen Institute for Quantum Science and Engineering and Department of Physics, Southern University of Science and Technology, Shenzhen 518055, China}
\affiliation{Center for Quantum Computing, Peng Cheng Laboratory, Shenzhen 518005, China}
\date{\today}

\begin{abstract}
One key feature of the multi-orbital superconducting \SRO~is the presence of nodal or near-nodal quasiparticle excitations revealed in a wide variety of experiments. Typically, a nodal gap structure in a two-dimensional model would be inconsistent with the chiral or helical p-wave interpretations. However, we demonstrate that true gap nodes may emerge along the $x$- and $y$-directions on the quasi-one-dimensional Fermi surfaces, if the multi-orbital chiral or helical p-wave pairings acquire peculiar forms wherein the $d_{xz}$ and $d_{yz}$ orbitals develop $k_y$- and $k_x$-like pairings, respectively. Spin-orbit coupling $\eta$ induces a near-nodal gap of order $\eta^2/W^2 \Delta_0$, where $\Delta_0$ is the gap amplitude and $W$ roughly the bandwidth. Provided the aforementioned pairing is predominant, the near-nodal gap structure is robust upon the inclusion of other multi-orbital pairings that share the same symmetries. In light of the recent experimental progresses, our proposal suggests that two-dimensional p-wave pairings may still be viable candidate ground states for \SRO. A near-nodal helical p-wave order, for example, may also be in line with the substantial drop in the NMR Knight shift under an in-plane magnetic field.
\end{abstract}

\maketitle
{\bf Introduction.} As an archetypal multi-orbital superconductor, \SRO~provides a fertile ground for studying unconventional Cooper pairing~\cite{Maeno:94,Mackenzie:03}. However, despite twenty-five years of intensive research, a coherent interpretation for its multiple key experimental signatures is still lacking~\cite{Rice:95,Kallin:09,Kallin:12,Maeno:12,Liu:15,Kallin:16,Mackenzie:17}. Earlier observations in support of the chiral p-wave ($p_x\pm ip_y$) order~\cite{Ishida:98,Duffy:00,Nelson:04,Luke:98,Xia:06} have been continuously challenged by various measurements. These include, but are not limited to, a substantial drop of the NMR Knight shift below $T_c$ under in-plane magnetic fields~\cite{Luo:19,Ishida:19}, and the indication of nodal quasiparticle excitations in electronic specific heat, nuclear spin relaxation $1/T_1$, ultrasound and thermal conductivity measurements~\cite{Nishizaki:00,Ishida:00,Deguchi:04,Lupien:01,Kittaka:18,Hassinger:17}.

The nodal behavior has stimulated a flurry of discussions. Foremost, a chiral p-wave pairing in a two-dimensional (2D) metal is generically fully gapped. The residual density of states have been attributed to anisotropic chiral p-wave pairing with deep gap minima (near-nodes) \cite{Zhitomirsky:01,Nomura:02,Raghu:10,Wang:13,Firmo:13,Scaffidi:14,Zhang:18,Wang:19,Roising:19}, or fundamentally different pairing symmetries, such as d-wave and highly anisotropic s-wave pairings, or a nematic p-wave ($p_x\pm p_y$)~\cite{Zhang:18,Huang:18,Huang:19a,Romer:19}. In 2D, the nematic p-wave phase is typically less favorable than the chiral phase. However, upon consideration of spin-orbit coupling (SOC) and out-of-plane pairing, a 3D nematic p-wave state, which respects time-reversal symmetry and has symmetry-protected point nodes, may be stabilized~\cite{Huang:18,Huang:19a}. There are also other proposals involving symmetry-protected or accidental horizontal line nodes, some of which are spin-triplet pairings and some are spin-singlet~\cite{Hasegawa:00,Annett:02,Ramires:19}.

Mindful of the presence of nodal or near-nodal excitations, we explore in this paper some peculiar forms of superconducting pairing made available by the multi-orbital nature of the Cooper pairing in this system. \SRO~has three bands, commonly denoted $\alpha$, $\beta$ and $\gamma$, crossing the Fermi level. They mainly derived from the three Ru $4d$ $t_{2g}$-orbitals~\cite{Damascelli:00,Bergemann:00}. We focus on the chiral and helical p-wave pairings in 2D models of \SRO, which have been frequently discussed in connection to this material. In particular, the helical p-wave pairing may be consistent with the observation of substantially suppressed NMR Knight shift in the presence of an in-plane magnetic field~\cite{Luo:19,Ishida:19}. We will show that, if the quasi-one-dimensional (1D) $d_{xz}$ and $d_{yz}$ orbitals respectively form $k_y$- and $k_x$-like pairings, near-nodal gaps develop along the high-symmetry x- and y-axis on the $\alpha$ and $\beta$ Fermi surfaces. In particular, these near-nodal gaps become true nodal points in the absence of SOC. So long as the aforementioned pairing remains dominant, inclusion of additional multi-orbital pairings that share the same symmetries does not qualitatively change the near-nodal gap structure.

The question about the primary superconducting orbital(s) in \SRO~has been a subject of intense theoretical debate~\cite{Agterberg:97,Raghu:10,Wang:13,Firmo:13,Huo:13,Scaffidi:14,Tsuchiizu:15,Huang:16,Gingras:18}. At the very lest, scanning tunneling spectroscopy seems to suggest that the $d_{xz}$ and $d_{yz}$ orbitals feature prominently in the superconducting state~\cite{Firmo:13}. Focusing only on the phenomenology, we shall refrain from discussing the microscopic mechanism for the particular form of the multi-orbital pairing. We shall also largely neglect the $d_{xy}$-orbital, but will briefly comment on its influence when necessary.

\begin{figure}
\includegraphics[width=0.45\textwidth]{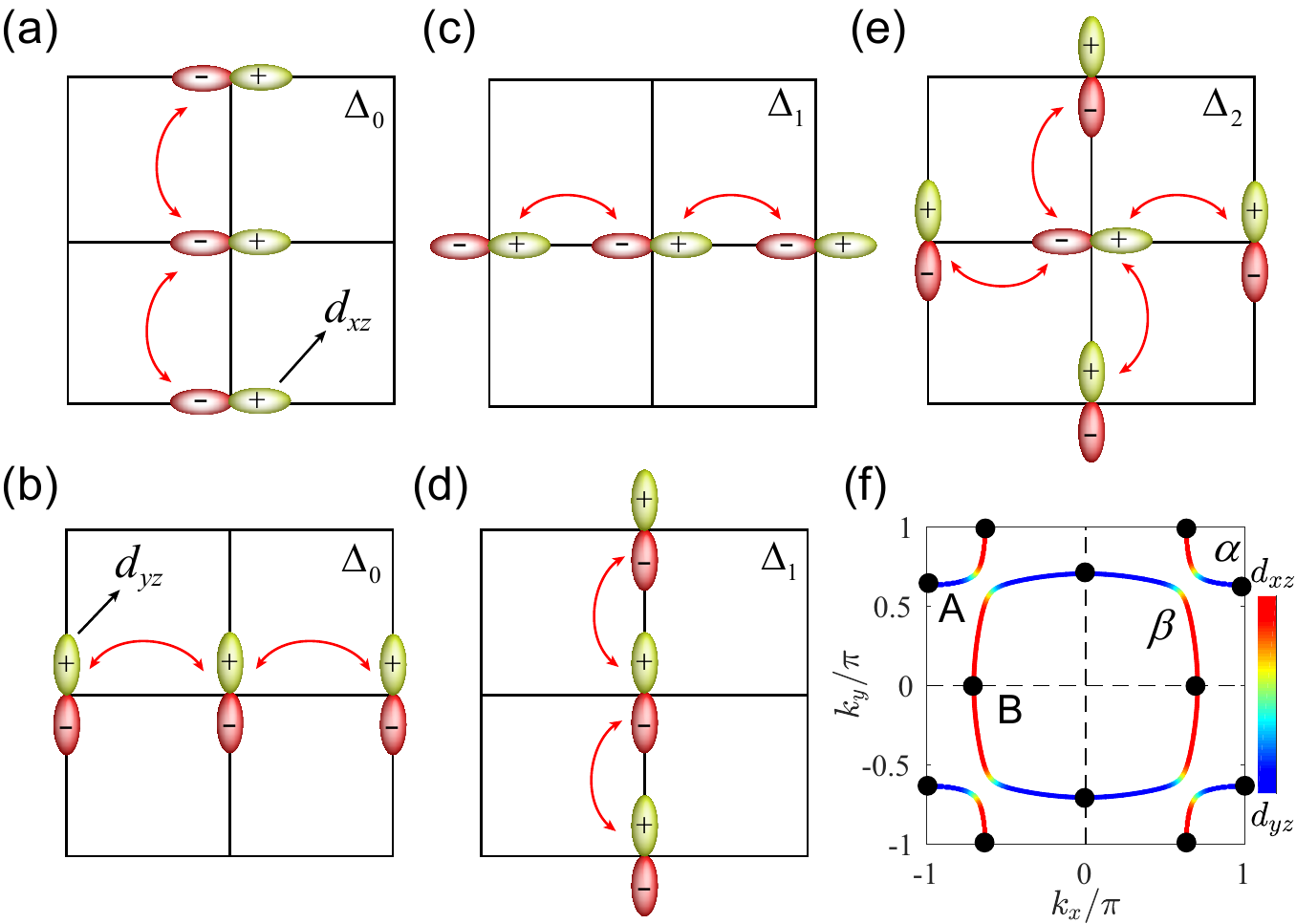}
\caption{(a)-(e) Top view sketch of the various forms of representative multi-orbital p-wave pairings in the two-orbital model. The pairings $\Delta_{i}$ ($i=0,1,2$) denote the corresponding $\hat{g}_{i}$ in Eq.~\ref{eqn2} for helical p-wave, and that of $\hat{f}_{i}$ in Eq.~\ref{eqn4} for chiral p-wave ($\Delta_i$ replaced by $\Delta_i^\prime$). Note that the spin indices are suppressed for clarity. The spatial form of the $\hat{g}_3$ pairing in the chiral state is also given by (e). In (a) and (b), the $d_{xz}$-orbital develops $\sin k_y$-like pairing and the $d_{yz}$-orbital forms $\sin k_x$-like pairing, respectively, and vice versa in (c) and (d). (e) Inter-orbital pairing which takes place on bonds in both directions. (f) Fermi surface with color-coding indicating the weight of the two orbitals on the Bloch bands. The highlighted points indicate the location of the (near)-nodal points.}
\label{fig:fig01}
\end{figure}

{\bf The model.} We first construct a two-orbital tight-binding Hamiltonian containing the $d_{xz}$ and $d_{yz}$-orbitals on a square lattice. In the spinor basis $(c_{x\bk\ua},c_{x\bk\da},c_{y\bk\ua},c_{y\bk\da})^T$ where the lower indices $x$ and $y$ label the $d_{xz}$ and $d_{yz}$-orbitals, respectively,
\begin{eqnarray}
H_{0\bk}&=&-2t_1(\cos k_x + \cos k_y)-\mu- 2t_2( \cos k_x -\cos k_y)\sigma_z \nonumber \\
&&+ 4t_3 \sin k_x\sin k_y\sigma_x + \eta\sigma_y \otimes s_z  \,.
\label{eq:H0}
\end{eqnarray}
Here $\sigma_i$ and $s_i$ with $i=x,y,z$ are the Pauli matrices operating on the respective orbital and spin degrees of freedom, $(t_1,t_2,t_3)$ designate the hopping integrals,  $\mu$ is the chemical potential and $\eta$ the magnitude of SOC. The two orbitals hybridize to form the quasi-1D $\alpha$ and $\beta$ bands. Setting $t_1+t_2=t$ and for a set of parameters widely used for \SRO, i.e. for $(t_1,t_2,t_3,\mu,\eta)=(0.55,0.45,0.1,1,0.1)t$, the two Fermi surfaces are plotted in Fig.~\ref{fig:fig01} (f). In the absence of SOC, the inter-orbital hopping, i.e. the term associated with $t_3$, vanishes along the $x$ and $y$ axis in the Brillouin zone (BZ). This will later become a crucial point of our proposal.

Refs. \onlinecite{Huang:19b,Ramires:19,Kaba:19} put forward extensive symmetry classifications of the multi-orbital pairing according to the $D_{4h}$ point group appropriate for \SRO. There are also related studies for iron-based superconductors~\cite{Dai:08,Zhou:08,Wan:08,Vafek:17,Cheung:19}. Although the details can be found in, e.g. Ref.~\cite{Huang:19b}, it is worth reiterating that symmetry operations, in addition to acting on the spatial and spin degrees of freedom, also transforms the orbital manifold. For instance, under a $C_4$ rotation, the orbitals transform as $d_{xz} \ra d_{yz}$ and $d_{yz} \ra -d_{xz}$. These multiple degrees of freedom provides a plethora of pairings not available in single-orbital systems.

{\bf Helical p-wave.} -- Recent NMR Knight shift measurement indicating a drop in the spin susceptibility in the superconducting state under in-plane magnetic fields, made the spin-triplet helical p-wave states -- which have their d-vector directed in-plane -- become appealing candidate superconducting states. There are in total four helical states, each belongs to one of the following irreducible representations (irrep): $A_{1u}$, $A_{2u}$, $B_{1u}$ or $B_{2u}$. Since our central argument applies to all of these distinct states, we take the $A_{1u}$ representation as an illustrative example. In single-orbital systems, this pairing has the simple basis function $k_x {\bf x}+k_y{\bf y}$, where the vectors ${\bf x}$ and ${\bf y}$ indicate the d-vector orientation of a spin-triplet pairing. However, the multi-orbital character introduces additional complexity. Represented in terms of the leading order lattice harmonics, the following basis functions all respect the $A_{1u}$ symmetry~\cite{Huang:19b},
\begin{eqnarray}
\hat{g}_0: && \sin k_x(\sigma_0 - \sigma_z)/2\otimes {\bf x}\cdot {\bf s}+ \sin k_y(\sigma_0 + \sigma_z)/2\otimes {\bf y}\cdot {\bf s} \nonumber \\
\hat{g}_1: && \sin k_x(\sigma_0 + \sigma_z)/2\otimes {\bf x}\cdot {\bf s}+ \sin k_y(\sigma_0 - \sigma_z)/2\otimes {\bf y}\cdot {\bf s} \nonumber \\
\hat{g}_2: && \sigma_x\otimes ( {\bf y}\sin k_x+  {\bf x}\sin k_y)\cdot{\bf s}
\label{eqn2}
\end{eqnarray}
All of these are spin-triplet pairings, and the corresponding superconducting gap function is given by $\hat{g}\cdot is_y$. Note that although we have, for simplicity, used only the simplest forms of lattice harmonics to describe the gap functions, realistic gaps could very well involve higher order terms. Some remarks are in order about the orbital structure of the individual terms, as illustrated in Fig.\ref{fig:fig01} (a)-(e). Both $\hat{g}_0$ and $\hat{g}_1$ are intra-orbital pairings. In $\hat{g}_0$, $d_{xz}$ and $d_{yz}$-orbitals form  $k_y$-like and $k_x$-like pairings, respectively, and vice versa in $\hat{g}_1$. The inter-orbital pairing $\hat{g}_2$ is even under orbital exchange, and hence is often referred to as orbital-triplet pairing.

It is also important to point out that the pairing is single-component (1D irrep), as the above pairing functions are inherently coupled by inter-orbital mixing and hence should in general coexist to form a single superconducting order parameter~\cite{Huang:19b}. Nevertheless, in subsequent discussions we assign to each basis an amplitude $\Delta_{i}$ ($i=0,1,2$) to describe their relative strength. The most general $A_{1u}$ pairing then takes the following form~\cite{Huang:19b},
\begin{eqnarray}
\hat{g}_\bk &=& \Delta_0 \hat{g}_{0\bk} + \Delta_1 \hat{g}_{1\bk}+\Delta_2 \hat{g}_{2\bk}
\end{eqnarray}
In consideration of the anisotropic dispersion relation of the quasi-1D orbitals, it would be tempting to assume that $\hat{g}_1$ dominates, as it has the Cooper pairing taking place along the primary hopping direction. In spite of this, the leading Cooper instability is determined, not only by the band structure, but also by the microscopic details of the bare electron-electron interactions and higher-order scattering processes. The pairing mechanism and the gap function are matters of ongoing discussions which do not seem to have immediate consensus in sight. For our purpose, we focus on a scenario in which the $\hat{g}_0$ pairing dominates over the other terms, i.e. $|\Delta_0| \gg |\Delta_{1,2}|$ (meanwhile, $|\Delta_0| \ll t$), which we argue will allow for near-nodal quasiparticle excitations.

\begin{figure}
\includegraphics[width=0.48\textwidth]{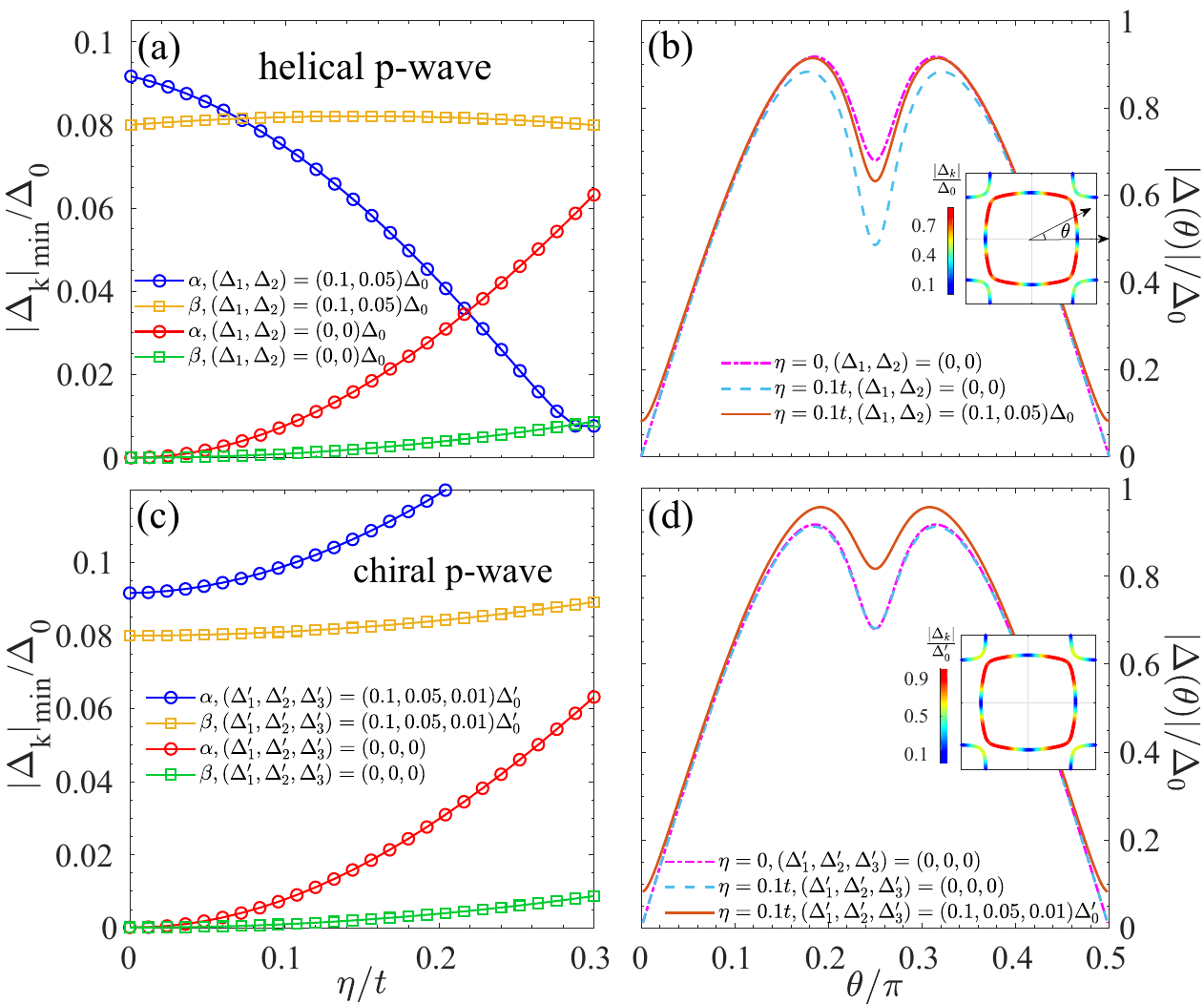}
\caption{Superconducting gap structure of the two-orbital model with various forms of multi-orbital pairings. The left panel, with (a) for helical p-wave in the $A_{1u}$ irrep and (c) for chiral p-wave, shows the near-nodal gap as a function of SOC at points A and B on the respective $\alpha$ and $\beta$ band Fermi surfaces in Fig.~\ref{fig:fig01} (f). In the right panel, (b) and (d) plot the azimuthal angular dependence of the gap on the $\beta$-band Fermi surface for helical and chiral states, respectively. The color-coded insets show the gap structure across the Fermi surfaces for $\eta=0.1t$.}
\label{fig:fig02}
\end{figure}

In the limiting case where the subleading pairings are both absent, i.e. $\Delta_{1,2}=0$, the BdG Hamiltonian returns four superconducting gap minima on each Fermi surface along the primary axis in the BZ, as indicated in Fig.~\ref{fig:fig01} (f). Most importantly, their magnitude falls to zero parametrically as a function of $\eta$ (Fig.~\ref{fig:fig02} (a)),
\begin{equation}
\Delta_\text{min} \simeq  \frac{a \eta^2}{W^2}|\Delta_0|
\label{eq:Gmin}
\end{equation}
where $W =2t+\mu$ is roughly an order of the bandwidth and the coefficient $a$ is another real constant determined by the tight-binding parameters which we comment on later. Given that $\eta/W \ll 1$ in \SRO, the robust near-nodal behavior is thus `imposed' by symmetry. This is our central message, which can be simply understood as follows. Let's take the limit $\eta=0$, under which circumstance the inter-orbital mixing vanishes along $k_x=0,\pi$ and $k_y=0,\pi$. For example, at point B in Fig.~\ref{fig:fig01} {\bf f}, the $\beta$-band is composed purely of the $d_{yz}$-orbital. As this orbital develops a $k_x$-like pairing, the gap on this band must vanish at this Fermi wavevector. Similar argument applies to all other highlighted points in Fig.~\ref{fig:fig01} {\bf f}. SOC induces orbital mixing and therefore opens a finite gap around the same momenta according to Eq.~\ref{eq:Gmin}. Note that at the value of $\eta \approx 0.1t$ appropriate for \SRO, the gap minima is more than two-orders of magnitude smaller than the gap amplitude. As the induced orbital mixing is more pronounced on the $\alpha$ band, the gap on this band typically exhibits stronger dependence on $\eta$, henceforth larger coefficient $a$ in Eq.~\ref{eq:Gmin}. Extending to 3D, the near-nodal points become vertical near-nodal lines.

Now consider adding subdominant pairings. It suffices to analyze the case with $\eta=0$. The intra-orbital pairing $\hat{g}_1$ generically gaps the entire spectrum. In fact, a gap of order $\Delta_1$ develops even when $\hat{g}_1$ appears alone. This again could be understood from the form of this intra-orbital pairing, which acquires the value $\Delta_1\sin k_\text{F}$ at the highlighted Fermi wavevector in Fig.~\ref{fig:fig01} {\bf f}. Further adding a weak inter-orbital pairing $\hat{g}_2$ influences the gaps quantitatively, but not qualitatively. Interestingly, when $\Delta_1=0$, the $\hat{g}_2$ pairing shifts the nodal points slightly away from the Fermi wavevectors. We note that, nodal excitations located off the Fermi surfaces are common occurrence in multi-orbital models and can be ascribed to the presence of interband pairing in an itinerant band-basis description. For completeness, Fig.~\ref{fig:fig02} also plots the gap minima and gap structure under finite-$\eta$.

Topologically, these near-nodal helical p-wave pairings are distinct from the traditionally discussed p-wave states where the $d_{xz}$ and $d_{yz}$ orbitals instead develop predominantly the $\Delta_1$ pairing. While the former is topologically trivial, the latter are characterized by nontrivial weak topological invariants and supports Majorana zero modes at crystalline defects and at the corners of certain sample geometries~\cite{Hughes:14,Benalcazar:14}. 

We briefly comment on the inclusion of the $d_{xy}$-orbital, which introduces a third, $\gamma$-band. Besides the conventional intra-orbital pairing ${\bf x}\sin k_x+ {\bf y}\sin k_y$, inter-orbital $A_{1u}$ pairings involving the $d_{xy}$-orbital are also possible. However, these additional pairings do not qualitatively affect the near-nodal excitations if the condition $\Delta_0\gg |\Delta_{1,2}|$ is satisfied on the $d_{xz}$ and $d_{yz}$ orbitals. This is because, firstly, on a square lattice the $d_{xy}$ orbital can only mix with the other orbitals through SOC, and secondly, such a mixing is maximum along the diagonals of the BZ and minimum along the $k_x$ and $k_y$ axis.

{\bf Chiral p-wave.} -- We now turn to the $E_{u}$ irrep, with which the chiral p-wave order is associated. Exactly the same argument for near-nodal gaps applies here. The following basis functions all belong to this representation~\cite{Huang:19b} (see Fig.~\ref{fig:fig01}),
\begin{eqnarray}
\hat{f}_0:&& \left[ \sin k_x (\sigma_0-\sigma_z)/2\otimes {\bf z}\cdot {\bf s}, \sin k_y (\sigma_0+\sigma_z)/2\otimes \bf{z}\cdot \bf{s} \right] \nonumber \\
\hat{f}_1:&& \left[ \sin k_x (\sigma_0+\sigma_z)/2\otimes {\bf z}\cdot {\bf s}, \sin k_y (\sigma_0-\sigma_z)/2\otimes \bf{z}\cdot \bf{s} \right] \nonumber \\
\hat{f}_2:&& \left(\sin k_y  \sigma_x\otimes {\bf z}\cdot {\bf s}, \sin k_x  \sigma_x\otimes {\bf z}\cdot {\bf s} \right) \nonumber \\
\hat{f}_3:&& \left( \sin k_x \cdot i\sigma_y, \sin k_y \cdot i\sigma_y \right)
\label{eqn4}
\end{eqnarray}
Each of the above terms has two components $\hat{f}_i\equiv (\hat{f}_i^x,\hat{f}_i^y)$. Notably, $\hat{f}_2$ has form factors $\sin k_x$ and $\sin k_y$ in reversed order. This ensures that all four of the $x$-components exhibit the same symmetry, and the same for the $y$-components. Among the four terms in (\ref{eqn4}), the first three are spin-triplet pairings, while the last one, $\hat{f}_3$, is a spin-singlet. Since spins are no longer good quantum numbers under finite SOC, spin-triplet and spin-singlet pairings are in general mixed~\cite{Huang:19b,Puetter:12,Veenstra:14}. Noteworthily, a mixture with sizable spin-singlet pairing could lead to a drop in the Knight shift.

As shown in Ref.~\cite{Huang:19b}, the combined effect of the SOC and inter-orbital hybridization results in a ubiquitous singlet-triplet entanglement. The two components of a general $E_u$ pairing reads,
\begin{eqnarray}
\hat{f}^x_\bk &=& \Delta^\prime_0 \hat{f}^x_{0\bk} + \Delta^\prime_1 \hat{f}^x_{1\bk}+\Delta^\prime_2 \hat{f}^x_{2\bk}+i\Delta^\prime_3\hat{f}^x_{3\bk} \nonumber \\
\hat{f}^y_\bk &=& \Delta^\prime_0 \hat{f}^y_{0\bk} + \Delta^\prime_1 \hat{f}^y_{1\bk}+\Delta^\prime_2 \hat{f}^y_{2\bk}-i\Delta^\prime_3\hat{f}^y_{3\bk}
\end{eqnarray}
where the prefactor $\pm i$ in front of the last terms can be traced back to the form of the SOC in Eq.~(\ref{eq:H0}). The new basis ($\hat{f}^x_\bk$, $\hat{f}^y_\bk$) forms a 2D $E_u$ irrep, and a chiral phase is stabilized if the two components develop an overall phase difference of $\pm \pi/2$. Most importantly, despite having a distinct spin configuration than $\hat{g}_0$, $\hat{f}_0$ is also in a form where the $d_{xz}$-orbital develops $k_y$-like pairing and the $d_{yz}$-orbital $k_x$-like pairing. Hence the $\hat{f}_0$ pairing, when appears alone, also supports nodal points in the absence of SOC. The influence of the SOC and the subdominant pairings $\hat{f}_{1,2,3}$ are also qualitatively similar to the case of the helical p-wave states, as shown in Fig.~\ref{fig:fig02} (b) and (d).

\begin{figure}
\includegraphics[width=0.45\textwidth]{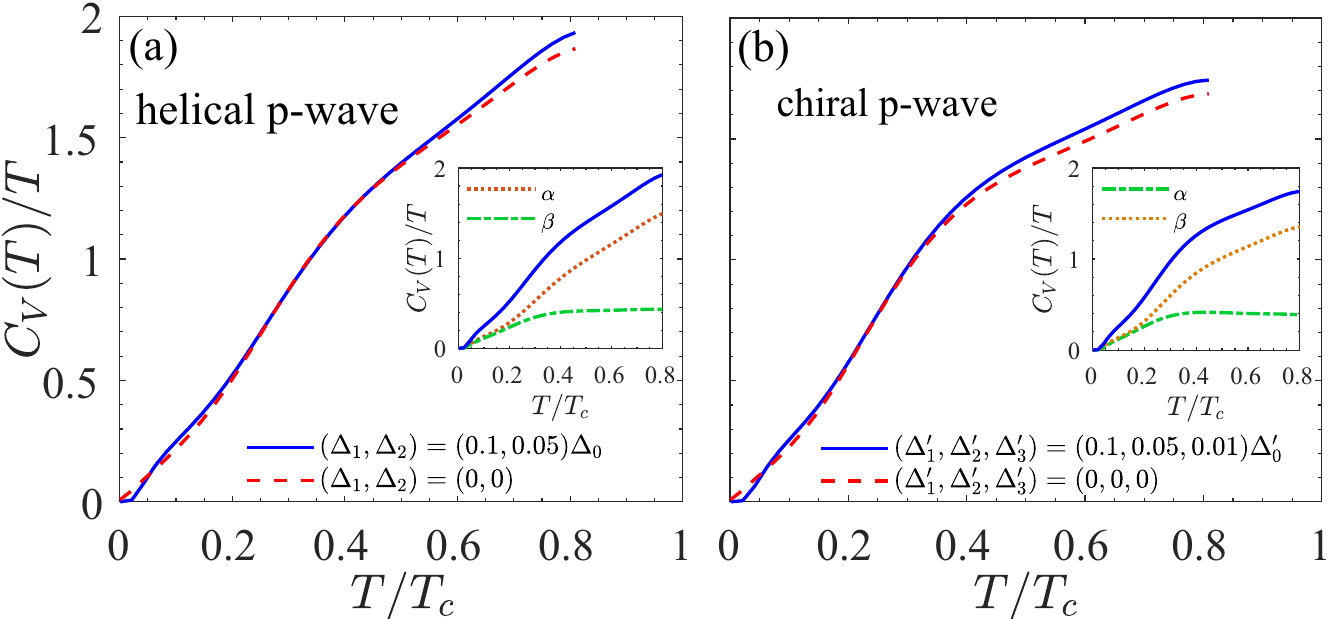}
\caption{Calculated specific heat of the multi-orbital model with $\eta=0.1t$ for (a) helical p-wave and for (b) chiral p-wave. The inset in each plot shows the contribution of the individual bands. }
\label{fig:fig03}
\end{figure}

{\bf Concluding remarks.}
The low-energy excitations naturally manifest in thermodynamic and transport measurements. In Fig.~\ref{fig:fig03}, we present the numerically evaluated specific heat for both states using a representative set of ${\Delta_i}$'s. Having neglected the $d_{xy}$-orbital and having used simplified gap functions, the results are not meant to be directly compared to the experimental data. However, the quasi-linear-$T$ behavior of $C_v/T$ over a wide low-temperature regime, reminiscent of the experimental observation~\cite{Nishizaki:00,Deguchi:04}, is robust against the microscopic details.

Throughout the work we have adopted the orbital-basis formulation. When transformed into the itinerant band-basis description, this formulation typically exhibits sizable interband Cooper pairing -- an unlikely scenario in the weak-coupling BCS theory. In spite of this, the near-nodal behavior remains robust even if the resultant interband pairing is purposely removed from our model.

There have been two other types of near-nodal points proposed within the p-wave framework of the quasi-2D models of \SRO. One of them emerges on the $\gamma$-band near the van Hove point where odd-parity pairing potentials must vanish~\cite{Nomura:02,Wang:13,Wang:19}. Here, the depth of the minima is constrained by the actual distance between the $\gamma$ Fermi surface and the van Hove point. The other type appears on the $\alpha$ or $\beta$ band at non-generic Fermi wavevectors and relies on the $d_{xz}$ and $d_{yz}$ orbitals developing Cooper pairings that additionally exhibit accidental zeros~\cite{Raghu:10,Firmo:13,Scaffidi:14,Zhang:18,Roising:19}. The near-nodal points along the high-symmetry axis on the quasi-1D bands proposed in the current work therefore represents a novel alternative possibility. They could be distinguished by quasiparticle interference measurements, although the \SRO~surface layer structural distortion may significantly complicate the experimental interpretations~\cite{Matzdorf:00,Veenstra:13}. 

As far as the (near)-nodal behavior is concerned, our proposal allows us to still think inside the conventional wisdom of 2D p-wave pairing for \SRO. In particular, a near-nodal helical p-wave pairing may simultaneously explain, at least, the suppressed Knight shift~\cite{Luo:19,Ishida:19} and in-plane $H_{c2}$~\cite{Deguchi:02,Rastovski:13,Kuhn:17}. However, these alone may not rule out other possibilities, such as d-wave or highly anisotropic s-wave pairing, or other more exotic forms of Cooper pairing~\cite{Huang:18,Huang:19a,Huang:19b,Ramires:19}. Any final identification must therefore be checked against the vast amount of other experimental realities, some of which are often ostentatiously at odds with each other~\cite{Maeno:12,Mackenzie:17}. Examples include the indications of multi-component pairing~\cite{Kidwingira:06,Saitoh:15,Wang:16,Anwar:17} in contrast with the lack of split superconducting transition under proper perturbations~\cite{Yonezawa:14,Li:19}, and the evidence of time-reversal symmetry breaking~\cite{Luke:98,Xia:06} against the absence of spontaneous surface current~\cite{Kirtley:07,Hicks:10,Curran:14}. We anticipate further efforts to reconcile our proposal with these key experimental observations.

{\bf Acknowledgement:} We would like to thank Yongkang Luo, Hong Yao and Yi Zhou for valuable discussions. This work is supported in part by a startup grant at the Southern University of Science and Technology.


\begin{thebibliography}{99}
\bibitem{Maeno:94} Y. Maeno, H. Hashimoto, {\it et al.}, Nature {\bf 372}, 532 (1994).
\bibitem{Mackenzie:03} A.P. Mackenzie and Y. Maeno, Rev. Mod. Phys. {\bf 75}, 657 (2003).
\bibitem{Rice:95} T.M. Rice and M. Sigrist, J. Phys.: Cond. Matt. {\bf 7}, L643 (1995).
\bibitem{Kallin:09} C. Kallin and A.J. Berlinsky, J. Phys. Condens. Matter {\bf 21}, 164210 (2009).
\bibitem{Kallin:12} C. Kallin, Rep. Prog. Phys. {\bf 75}, 042501 (2012).
\bibitem{Maeno:12} Y. Maeno, S. Kittaka, T. Nomura, S. Yonezawa, and K. Ishida, J. Phys. Soc. Jpn. {\bf 81}, 011009 (2012).
\bibitem{Liu:15} Y. Liu and Z.Q. Mao, Physica C: Superconductivity and its Application, {\bf 514}, 339 (2015).
\bibitem{Kallin:16} C. Kallin and A.J. Berlinsky, Rep. Prog. Phys. {\bf 79}, 054502 (2016).
\bibitem{Mackenzie:17} A. P. Mackenzie, T. Scaffidi, C. W. Hicks and Y. Maeno, NPJ Quantum Materials {\bf 2}, 40 (2017).
\bibitem{Ishida:98} K. Ishida {\it et al.}, Nature (London) {\bf 396}, 658 (1998).
\bibitem{Duffy:00} J. A. Duffy, S. M. Hayden, Y. Maeno, Z. Mao, J. Kulda, and G.J. McIntyre, Phys. Rev. Lett. {\bf 85}, 5412 (2000).
\bibitem{Nelson:04} K.D. Nelson, Z.Q. Mao, Y. Maeno, and Y. Liu, Science {\bf 306}, 1151 (2004).
\bibitem{Luke:98} G.M. Luke {\it et al.}, Nature (London) {\bf 394}, 558 (1998).
\bibitem{Xia:06} J. Xia, Y. Maeno, P.T. Beyersdorf, M.M. Fejer, and A. Kapitulnik, Phys. Rev. Lett. {\bf 97}, 167002 (2006).
\bibitem{Luo:19} A. Pustogow, Y. Luo, A. Chronister, {\it et al.}, arXiv:1904.00047 (2019).
\bibitem{Ishida:19} K. Ishida, Masahiro Manago and Y. Maeno, arXiv:1907.12236.
\bibitem{Nishizaki:00} S. NishiZaki, Y. Maeno, and Z.Q. Mao, J. Phys. Soc. Jpn. {\bf 69}, 572 (2000).
\bibitem{Ishida:00} K. Ishida, H. Mukuda, Y. Kitaoka, Z. Q. Mao, Y. Mori, and Y. Maeno, Phys. Rev. Lett. {\bf 84}, 5387 (2000).
\bibitem{Deguchi:04} K. Deguchi, Z. Q. Mao, H. Yaguchi, and Y. Maeno, Phys. Rev. Lett. {\bf 92}, 047002 (2004).
\bibitem{Lupien:01} C. Lupien, W. A. MacFarlane, C. Proust, L. Taillefer, Z. Q. Mao, and Y. Maeno, Phys. Rev. Lett. {\bf 86}, 5986 (2001).
\bibitem{Kittaka:18} S. Kittaka, S. Nakamura, T. Sakakibara, N. Kikugawa, T. Terashima, S. Uji, D.A. Sokolov, A. P. Mackenzie, K. Irie, Y. Tsutsumi, K. Suzuki, K. Machida, J. Phys. Soc. Jpn. 87, 093703 (2018).
\bibitem{Hassinger:17} E. Hassinger, P. Bourgeois-Hope, H. Taniguchi, S. RenedeCotret, G. Grissonnanche, M. S. Anwar, Y. Maeno, N. Doiron-Leyraud, and L. Taillefer, Phys. Rev. X {\bf 7}, 011032 (2017).
\bibitem{Zhitomirsky:01} M. E. Zhitomirsky and T.M. Rice, Phys. Rev. Lett. {\bf 87}, 057001 (2001).
\bibitem{Nomura:02} T. Nomura and K. Yamada: J. Phys. Soc. Jpn. {\bf 71}, 404 (2002).
\bibitem{Raghu:10} S. Raghu, A. Kapitulnik, and S. A. Kivelson, Phys. Rev. Lett. 105, 136401 (2010).
\bibitem{Firmo:13} I. A. Firmo, S. Lederer, C. Lupien, A. P. Mackenzie, J. C. Davis, and S. A. Kivelson, Phys. Rev. B {\bf 88}, 134521 (2013).
\bibitem{Wang:13} Q.H. Wang, C. Platt, Y. Yang, C. Honerkamp, F.C. Zhang, W. Hanke, T.M. Rice and R. Thomale, Europhys. Lett. {\bf 104}, 17013 (2013).
\bibitem{Scaffidi:14} T. Scaffidi, J. C. Romers, and S. H. Simon, Phys. Rev. B {\bf 89}, 220510(R) (2014).
\bibitem{Zhang:18} L-D. Zhang, W. Huang, F. Yang and H. Yao, Phys. Rev. B {\bf 97}, 060510(R) (2018).
\bibitem{Wang:19} W-S. Wang, C-C. Zhang, F-C. Zhang, Q-H. Wang, Phys. Rev. Lett. {\bf 122}, 027002 (2019).
\bibitem{Roising:19} H.S. R\o ising, T. Scaffidi, F. Flicker, G.F. Lange, and S.H. Simon, arXiv:1907.09485.
\bibitem{Huang:18} W. Huang and H. Yao, Phys. Rev. Lett. {\bf 121}, 157002 (2018).
\bibitem{Huang:19a} W. Huang, Y. Zhou, and H. Yao, arXiv:1901.07041.
\bibitem{Romer:19} A. T. R\o mer, D. D. Scherer, I. M. Eremin, P. J. Hirschfeld, B. M. Andersen, arXiv:1905.04782.
\bibitem{Hasegawa:00} Y. Hasegawa, K. Machida and M. Ozaki, J. Phys. Soc. Jpn. {\bf 69}, 336 (2000).
\bibitem{Annett:02} J. F. Annett, G. Litak, B.L. Gy\"orffy, and K.I. Wysokinski, Phys. Rev. B {\bf 66}, 134514 (2002).
\bibitem{Ramires:19} A. Ramires and M. Sigrist, arXiv:1905.01288.
\bibitem{Damascelli:00} A. Damascelli, {\it et. al.}, Phys. Rev. Lett. {\bf 85}, 5194 (2000).
\bibitem{Bergemann:00} C. Bergemann, S.R. Julian, A.P. Mackenzie, S. NishiZaki and Y. Maeno, Phys. Rev. Lett. {\bf 84}, 2662 (2000).
\bibitem{Agterberg:97} D. F. Agterberg, T. M. Rice, and M. Sigrist, Phys. Rev. Lett. {\bf 78}, 3374 (1997).
\bibitem{Huo:13} J.W. Huo, T.M. Rice and F.C. Zhang, Phys. Rev. Lett. {\bf 110}, 167003 (2013).
\bibitem{Huang:16} W. Huang, T. Scaffidi, M. Sigrist, and C. Kallin, {\bf 94}, 064508 (2016).
\bibitem{Tsuchiizu:15}  M. Tsuchiizu, Y. Yamakawa, S. Onari, Y. Ohno, and H. Kontani, Phys. Rev. B {\bf 91}, 155103 (2015).
\bibitem{Gingras:18} O. Gingras, R. Nourafkan, A.-M. S. Tremblay, M. C\`{o}t\'{e}, arXiv:1808.02527.
\bibitem{Huang:19b} W. Huang, Y. Zhou, and H. Yao, arXiv:1905.03523.
\bibitem{Kaba:19} S.-O. Kaba, D. S\'en\'echal, arXiv:1905.10467.
\bibitem{Dai:08} X. Dai, Z. Fang, Y. Zhou, and F.-C. Zhang, Phys. Rev. Lett. {\bf 101}, 057008 (2008).
\bibitem{Zhou:08} Y. Zhou, W.Q. Chen, and F.C. Zhang, Phys. Rev. B {\bf 78}, 064514 (2008).
\bibitem{Wan:08} Y. Wan and Q.H. Wang, Europhys. Lett. 85 57007 (2009).
\bibitem{Vafek:17} O. Vafek and A.V. Chubukov, Phys. Rev. Lett. {\bf 118}, 087003 (2017).
\bibitem{Cheung:19} A.K.C. Cheung and D.F. Agterberg, Phys. Rev. B {\bf 99}, 024516 (2019).
\bibitem{Hughes:14} T.L. Hughes, H. Yao, and X-L. Qi, Phys. Rev. B {\bf 90}, 235123 (2014). 
\bibitem{Benalcazar:14} W. A. Benalcazar, J. C. Teo, and T. L. Hughes, Phys. Rev. B 89, 224503 (2014).
\bibitem{Puetter:12} C.M. Puetter and H-Y. Kee, EPL, {\bf 98}, 27010 (2012).
\bibitem{Veenstra:14} C. N. Veenstra, Z.-H. Zhu, M. Raichle, {\it et al.}, Phys. Rev. Lett. {\bf 112}, 127002 (2014).
\bibitem{Matzdorf:00} R. Matzdorf, Z. Fang, Ismail, J. Zhang, T. Kimura, Y. Tokura, K. Terakura, and E.W. Plummer, Science {\bf 289}, 746 (2000).  
\bibitem{Veenstra:13} C.N. Veenstra, Z.-H. Zhu, B. Ludbrook, M. Capsoni, G. Levy, A. Nicolaou, J. A. Rosen, R. Comin, S. Kittaka, Y. Maeno, I. S. Elfimov, A. Damascelli, Phys. Rev. Lett. {\bf 110}, 097004 (2013). 
\bibitem{Deguchi:02} K. Deguchi, M.A. Tanatar, Z.Q. Mao, T. Ishiguro, and Y. Maeno, J. Phys. Jpn. Soc. {\bf 71}, 2839 (2002).
\bibitem{Rastovski:13} C. Rastovski, C.D. Dewhurst, W.J. Gannon, {\it et al.}, Phys. Rev. Lett. {\bf 111}, 087003 (2013).
\bibitem{Kuhn:17} S.J. Kuhn, W. Morgenlander, E.R. Louden, {\it et al.}, Phys. Rev. B {\bf 96}, 174507 (2017).
\bibitem{Kidwingira:06} F. Kidwingira, J.D. Strand, D.J. Van Harlingen, Y. Maeno, Science {\bf 314}, 1276 (2006).
\bibitem{Saitoh:15} K. Saitoh, S. Kashiwaya, {\it et al.}, Phys. Rev. B {\bf 92}, 100504(R) (2015).
\bibitem{Wang:16} H. Wang, J. Luo, W. Lou, J. Wei, {\it et al.}, New J. Phys. {\bf 19}, 053001 (2017).
\bibitem{Anwar:17} M.S. Anwar, R. Ishiguro, {\it et al.}, Phys. Rev. B {\bf 95}, 224509 (2017).

\bibitem{Li:19} Y.-S. Li, N. Kikugawa, D.A. Sokolov, F. Jerzembeck, A.S. Gibbs, Y. Maeno, C.W. Hicks, M. Nicklas, A.P. Mackenzie, arXiv:1906.07597.
\bibitem{Yonezawa:14} S. Yonezawa, T. Kajikawa, Y. Maeno, J. Phys. Soc. Jpn. {\bf 83}, 083706 (2014).
\bibitem{Kirtley:07} J.R. Kirtley, C. Kallin, C.W. Hicks, E.-A. Kim, Y. Liu, K.A. Moler, Y. Maeno, K.D. Nelson, Phys. Rev. B {\bf 76}, 014526 (2007).
\bibitem{Hicks:10} C.W. Hicks, J.R. Kirtley, T.M. Lippman, N.C. Koshnick, {\it et al.}, Phys. Rev. B {\bf 81}, 214501 (2010).
\bibitem{Curran:14} P.J. Curran, S.J. Bending, W.M. Desoky, A.S. Gibbs, S.L. Lee, A.P. Mackenzie, Phys. Rev. B {\bf 89}, 144504 (2014).

\end{thebibliography}
\end{document}